\documentclass[proof]{WileyASNA-v1}

\articletype{Article Type}%

\received{29 November 2019}
\revised{x}
\accepted{x}

\begin{document}

\title{Charge Exchange, from the sky to the laboratory: a method to determine state-selective cross sections for improved modeling}

\author[1]{Gabriele L. Betancourt-Martinez*}
\author[2,3]{Renata S. Cumbee}
\author[2]{Maurice A. Leutenegger}

\address[1]{\orgname{IRAP/CNRS}, \orgaddress{\state{Toulouse}, \country{France}}}
\address[2]{\orgname{NASA/GSFC}, \orgaddress{\state{Greenbelt}, \country{USA}}}
\address[3]{\orgname{University of Maryland, College Park}, \orgaddress{\state{College Park}, \country{USA}}}

\corres{*G. L. Betancourt-Martinez \email{gabriele.betancourt@irap.omp.eu}}

\abstract{Charge exchange (CX) is a semi-resonant recombination process that can lead to spectral line emission in the X-ray band. It occurs in nearly any environment where hot plasma and cold gas interact: in the solar system, in comets and planetary atmospheres, and likely astrophysically, in, for example, supernova remnants and galaxy clusters. It also contributes to the soft X-ray background. Accurate spectral modeling of CX is thus critical to properly interpreting our astrophysical observations, but commonly-used CX models in popular spectral fitting packages often rely on scaling equations, and may not accurately describe observations or laboratory measurements. This paper introduces a method that can be applied to high-resolution CX spectra to directly extract state-selective CX cross sections for electron capture, a key parameter for properly simulating the resulting CX spectrum.}


\maketitle

\footnotetext{\textbf{Abbreviations:} CX, charge exchange; EBIT, electron beam ion trap; ECS, EBIT calorimeter spectrometer; FAC, flexible atomic code; MCLZ, multichannel Landau-Zener; SEC, single electron capture; SWCX, solar wind charge exchange}

\section{Charge Exchange Observations}\label{sec1}

Charge Exchange (CX) is a semi-resonant process in which one or more electrons are transferred from a neutral atom or molecule into an excited state of an ion. Line emission, often in the X-ray band, is then produced when the excited electron radiatively de-excites to the ground state. CX X-ray emission from ions in the solar wind (solar wind CX, or SWCX) has been observed around comets \citep{1996Sci...274..205L, 1997GeoRL..24..105C} and planetary atmospheres \citep{2002Natur.415.1000G, 2002A&A...394.1119D,2012AN....333..324D}, and is a non-negligible contribution to the local X-ray background \citep{2004ApJ...610.1182S, 2014Natur.512..171G}. Notably, the spectral signature of SWCX is temporally variable at multiple scales, depending on the source of the neutrals; heliospheric CX, varying on the timescale of days to weeks, is particularly difficult to identify and subtract in observations. Astrophysically, there are hints of CX in objects such as clusters of galaxies \citep{2015MNRAS.453.2480W, 2017ApJ...837L..15A}, starbust galaxies \citep{2007PASJ...59S.269T}, and supernova remnants \citep{2015MNRAS.449.1340R}. CX is thus relevant to our current observations from, e.g., \textit{XMM-Newton} and \textit{Chandra}, as well as from future missions such as \textit{XRISM} and \textit{Athena}. The spectral signatures of CX are highly dependent on---and thus diagnostic of---the relative velocity between the ion and neutral, as well as the specific ion and neutral species and densities. In order to properly interpret these diagnostics, we rely on accurate CX models in the spectral synthesis codes we use.  

\section{Charge Exchange Models in \textsc{xspec} and \textsc{spex}}\label{sec2}

Two commonly used spectral fitting packages are \textsc{xspec} and \textsc{spex}, both of which include models for CX. \textsc{acx} \citep{2014ApJ...787...77S}, a standalone package that can be used in the \textsc{xspec} spectral fitting package, recently has been expanded to two versions: \textsc{acx1} is the default model usable with \textsc{xspec}, and \textsc{acx2} can be used with the \textsc{PyXspec} library. \textsc{acx1} employs scaling equations based on \citet{1985PhR...117..265J} that use the charge of the ion and the ionization potential of the neutral to determine the initial $n$ and $l$ distributions of the captured electrons after the transfer. The \textit{model} parameter in the model allows the user to change between scaling equations, and thus different distributions. In contrast to using purely empirical equations, \textsc{acx2} uses calculated $nl$-selective, velocity-dependent cross sections from the \textit{Kronos} database \citep{2016ApJS..224...31M} for certain ions. In the absence of \textit{Kronos} cross sections, it reverts back to the scaling equations. For both \textsc{acx1} and \textsc{acx2}, after determining the excited state of the ion, the subsequent radiative cascade is calculated using atomic parameters from AtomDB \citep{2011nlaw.confC...2F} to simulate the spectrum\textsc{spex-cx} \citep{2016A&A...588A..52G} adopts $n$- or $nl$-selective cross sections collected from the literature. To fill in incomplete data, \textsc{spex-cx} falls back on derived scaling laws for the $n$ and $l$ distribution (see \citet{2016A&A...588A..52G} for details). It then calculates the radiative cascade using the \textsc{spex} atomic database \citep{1996uxsa.conf..411K}. Both \textsc{spex-cx} and \textsc{acx} models assume purely single electron capture (SEC), and a neutral partner of H and He (\textsc{acx}) or only H (\textsc{spex-cx}), and only \textsc{acx2} and \textsc{spex-cx} allow the user to specify a ion/neutral collision velocity. 

\begin{figure}
 \begin{center}
 \includegraphics[scale=0.35]{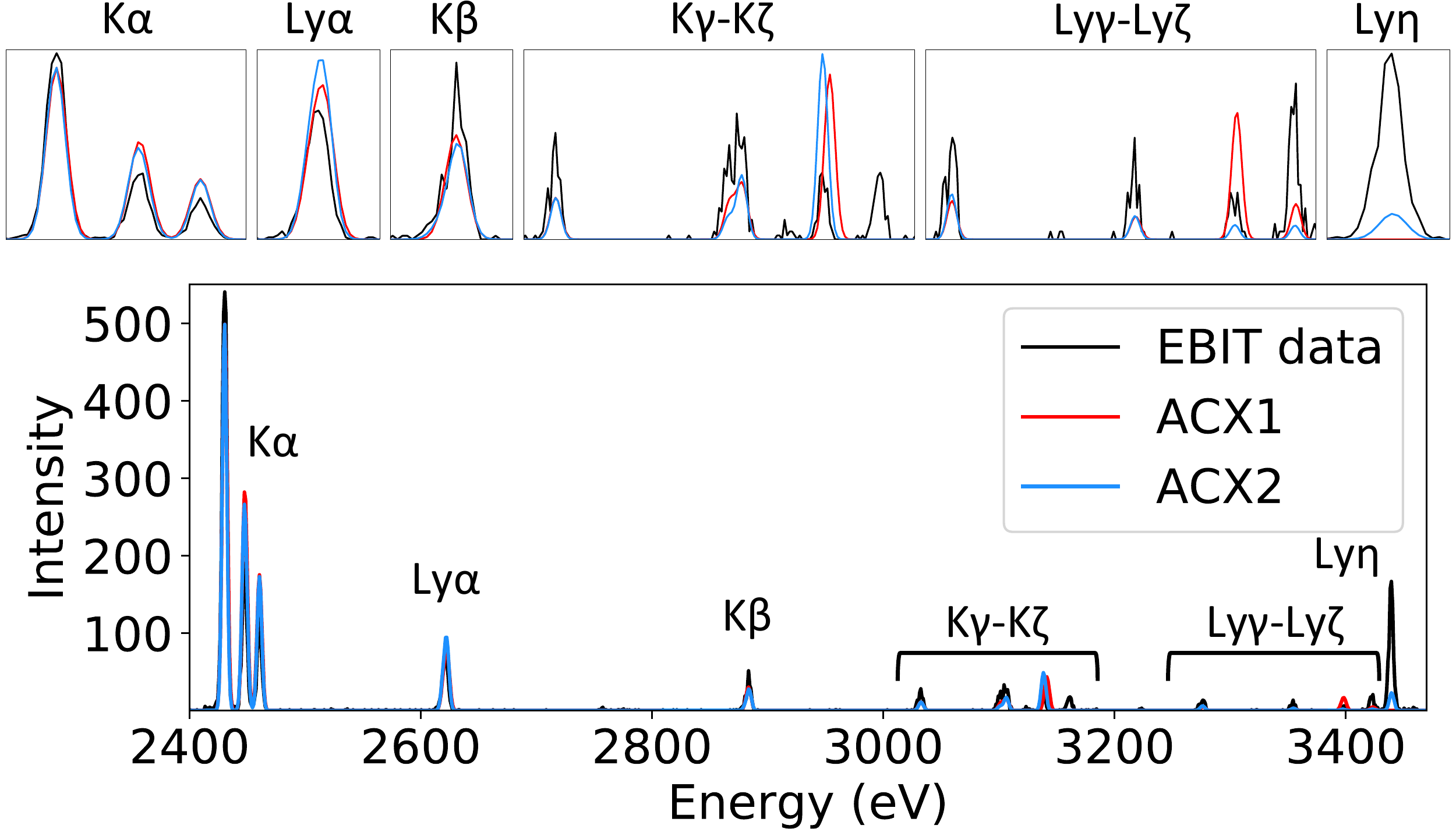}
 \end{center}
 \renewcommand{\baselinestretch}{1}
\small\normalsize
\begin{quote}
\caption[]{Laboratory CX data (black) of H- and He-like S undergoing CX with neutral He, compared to the corresponding \textsc{acx1} and \textsc{acx2} models. The full spectrum is shown on the bottom, with magnifications of the lines shown above. Complete line labels are shown in Figures \ref{fig:S_model_data} and \ref{fig:simspec}.}
\label{fig:acxcompare}
\end{quote}
\end{figure}

While \textsc{acx} and \textsc{spex-cx} use different databases for the radiative cascade, it is the initial $nl$ distribution of captured electrons that has the dominant impact on the resulting spectrum---in particular, the $l$ distribution, which is highly velocity-dependent \citep{2000PhRvL..85.5090B, 2014PhRvA..90e2723B, 2016ApJS..224...31M}. The \textsc{acx} and \textsc{spex-cx} models provide good fits to certain observations of SWCX \citep{2014ApJ...787...77S, 2016A&A...588A..52G}, but they fail to reproduce some laboratory measurements of CX in the low-velocity regime  \citep{2018ApJ...868L..17B}. This is problematic because these experimental benchmarks are performed in order to validate theory. An example of this discrepancy is shown in Figure \ref{fig:acxcompare}, which depicts a laboratory spectrum of K-shell S undergoing CX with neutral He, compared to \textsc{acx} models. 

\graphicspath{{images/}}

The greatest need for more accurate CX modeling is improved and expanded theoretical state-selective (at least $nl$-resolved) cross sections \citep{2014AAMOP..63..271S, 2007RvMP...79...79K, 2019BAAS...51c.337B} as a function of collision velocity. These must be rigorously benchmarked to experiments. To address this need, we present a method that can be used in parallel with or in the absence of these theoretical values, which extracts these key values directly from high-resolution laboratory, observational, or simulated spectra. 

\section{A Method to Extract State-Selective Cross Sections}\label{method}

In this paper, we describe the overall method, apply it to the simple case of a laboratory K-shell spectrum, and compare our results to existing theory. A subsequent paper will apply the method to L-shell ions.  

This procedure involves four basic steps, which will be described in more detail in the following paragraphs. First, using the Flexible Atomic Code (\textsc{fac}, \citet{2008CaJPh..86..675G}) we calculate the atomic structure of several configurations of the projectile ion(s) just after SEC, i.e., the base ion with the addition of one excited electron, where we choose many possible excited states. Second, we use the atomic data to perform a radiative cascade following de-excitation from each of these states, creating a spectrum for each one. Third, from this set of spectra, we select a subset to form a spectral basis set which we add together into a model to fit our laboratory data. Finally, we run our model, which uses this spectral basis set in a least-squares minimization procedure to calculate the relative weight to be applied to each spectrum to create a best fit to the data. As each spectrum in the basis set is representative of electron capture into one specific quantum state, the weight assigned to each spectrum in the basis set after the fit corresponds to the relative cross section of electron capture into that quantum state. 

To elaborate on each of these steps with a concrete example, we consider a laboratory CX spectrum obtained with the EBIT-I Electron Beam Ion Trap (EBIT) \citep{1989AIPC..188...82L} at the Lawrence Livermore National Laboratory using the EBIT Microcalorimeter Spectrometer (ECS, \citet{2008JLTP..151.1061P}) which has $\sim$4 eV resolution at 6 keV. This data was initially published in \citet{2014PhRvA..90e2723B}, and includes bare and H-like sulphur (S\textsuperscript{16+}, S\textsuperscript{15+}) that, after SEC from neutral He, become H- and He-like (S\textsuperscript{15+}, S\textsuperscript{14+}). The spectrum is shown in Figures \ref{fig:acxcompare} and \ref{fig:S_model_data}. Our first step in the method is to consider two different species of S ion: S\textsuperscript{15+} with one excited electron ($nljJ^*$), and S\textsuperscript{14+} with one electron in its ground state and the second electron in an excited state ($1s_{1/2}^1nljJ^*$). We consider many quantum states for the excited electron, going as high in energy as $n=15$. This is done with \textsc{fac}, assuming $jj$-coupling. 

Next, we use the aforementioned atomic data to calculate a radiative cascade matrix between all calculated energy levels in these excited H- and He-like ions. At this stage, we begin to narrow the subset of excited states that we consider by choosing a range of $n$ values that are likely to be most relevant to our experimental spectra. We do this by examining the experimental spectrum to determine the highest energy spectral line with significant flux, then conservatively choosing an $n$\textsubscript{min} and $n$\textsubscript{max} range that bounds this primary capture channel, where $n$\textsubscript{max}-- $n$\textsubscript{min}$>5$. For example, our S spectrum shows a strong high-$n$ line that corresponds to the $n=8\rightarrow 1$ transition near 3.44 keV (see Figure \ref{fig:acxcompare}), so we limit our considered configurations to those with an excited electron in $n=4-10$. We include all possible $l$ and $j$ values, as well as coupled $J$ values for He-like ions, within this range. We input this narrowed set of excited configurations into the cascade matrix to generate the resulting spectra. Each spectrum is thus the result from cascade from a given excited state. 

For our example case, we now have a set of spectra that corresponds to H- and He-like S ions that have just undergone CX with SEC (into $n=4-10$) and have radiatively de-excited to the ground state. This is a large number of spectra---on the order of hundreds---which is more than should be used to fit a spectrum with only about 15 lines. It is thus necessary to reduce the number of spectra that will make up a basis set for our fitting code. Choosing the spectra to make up a basis set may be accomplished in various ways; the spectra in the basis sets used in this work were chosen by visual inspection based on their ability to produce lines present in our experimental spectra. They were generally limited to low values of $l$ to reflect the low collision velocities that occur in EBIT experiments, and are weighted more heavily towards the observed $n$\textsubscript{max} from the data. We repeated this process with different spectral basis sets to determine how this changed both the model spectrum generated and the resulting cross sections. There are also degeneracy effects to consider: cascades following several different excited electron configurations might lead to very similar spectra. Thus each element in our chosen basis set may not be representative of a single $n,l,j,J$-resolved state, but a sum of several. Our follow-up paper will introduce various methods to achieve this reduction in spectra, in particular to reduce biases by taking a more randomized approach. 

With a chosen spectral basis set, the fourth and final stage of our method is then inputting our basis set into our model, which we fit to our experimental data using the least-squares minimization Python package \textsc{lmfit}. The parameters of our model are the capture states themselves (i.e., the spectra in our basis set), the Gaussian broadening of the spectral lines, and the overall weight of each spectrum in the basis set. The basis set is fixed, the Gaussian broadening for all spectral lines is tied together, and we allow the overall weighting to vary. For example, for a dummy basis set with three simulated spectra A, B, and C, the best fit to the data can be represented as:

\begin{equation}
\text{Best fit  = (Spec A*W1) + (Spec B*W2) + (Spec C*W3)},
\end{equation}
where W1, W2, and W3 correspond to the weights assigned to spectra A, B, and C, respectively, after the fit. These weights correspond to the relative cross section of the $n,l,j,J$-resolved capture state that lead to that spectrum after radiative decay. 

\section{Method Performance}\label{perfo}

\begin{figure}
 \begin{center}
 \includegraphics[scale=0.55]{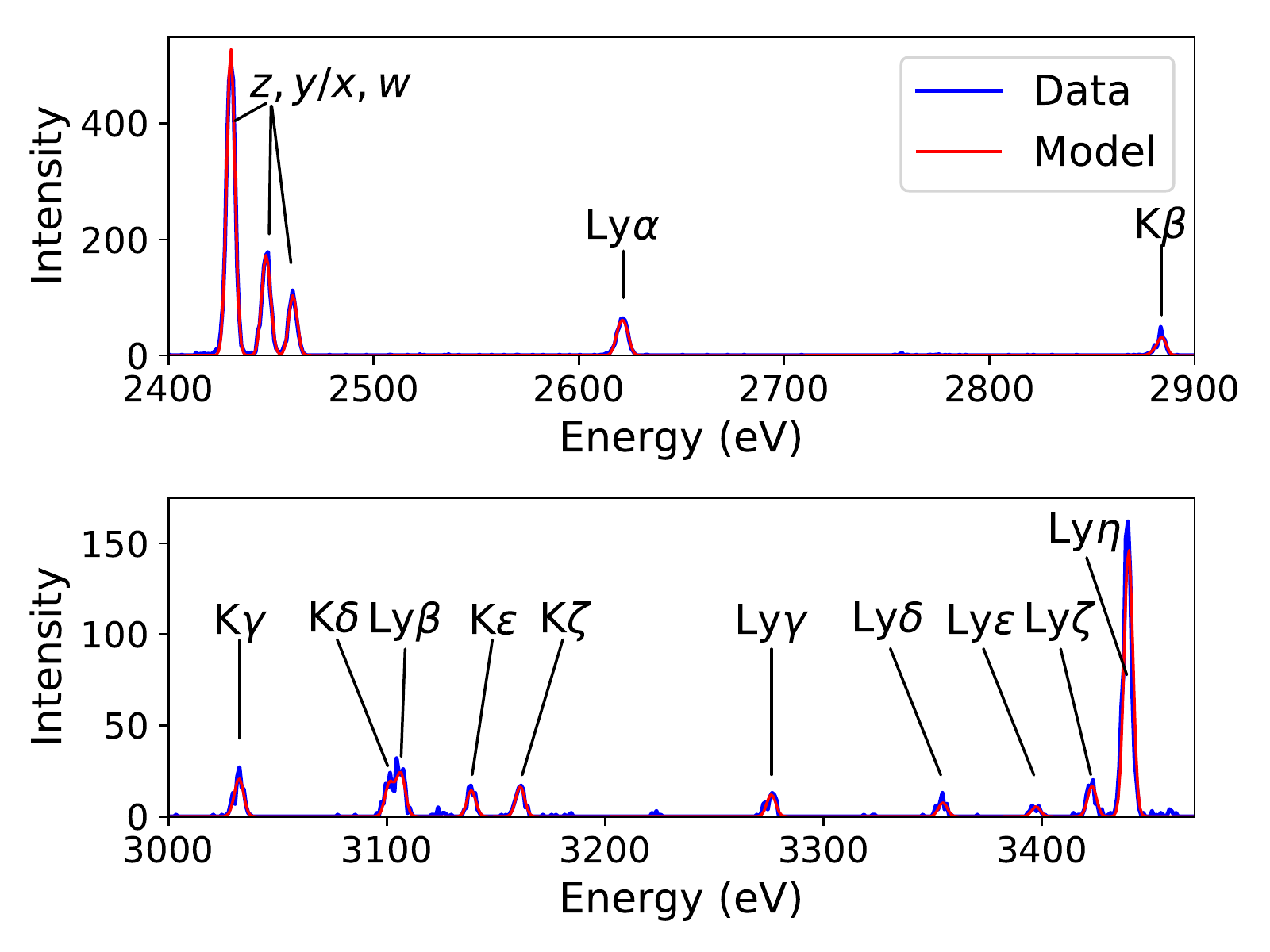}
 \end{center}
 \renewcommand{\baselinestretch}{1}
\small\normalsize
\begin{quote}
\caption[Model fit to the data for H- and He-like S+He CX.]{Model fit (red) to the data (blue) for H- and He-like S+He CX.}
\label{fig:S_model_data}
\end{quote}
\end{figure}

The spectrum generated with the best fit to the S+He CX data is shown in Figure \ref{fig:S_model_data}. In general, the fit is very good: though the He-like forbidden line near 2400 eV is slightly over-predicted and the H-like Ly-$\eta$ line near 3450 eV is slightly under-predicted, most other lines are well-described by the model. 

\graphicspath{{images/}}

\begin{figure}
 \begin{center}
 \includegraphics[scale=0.55]{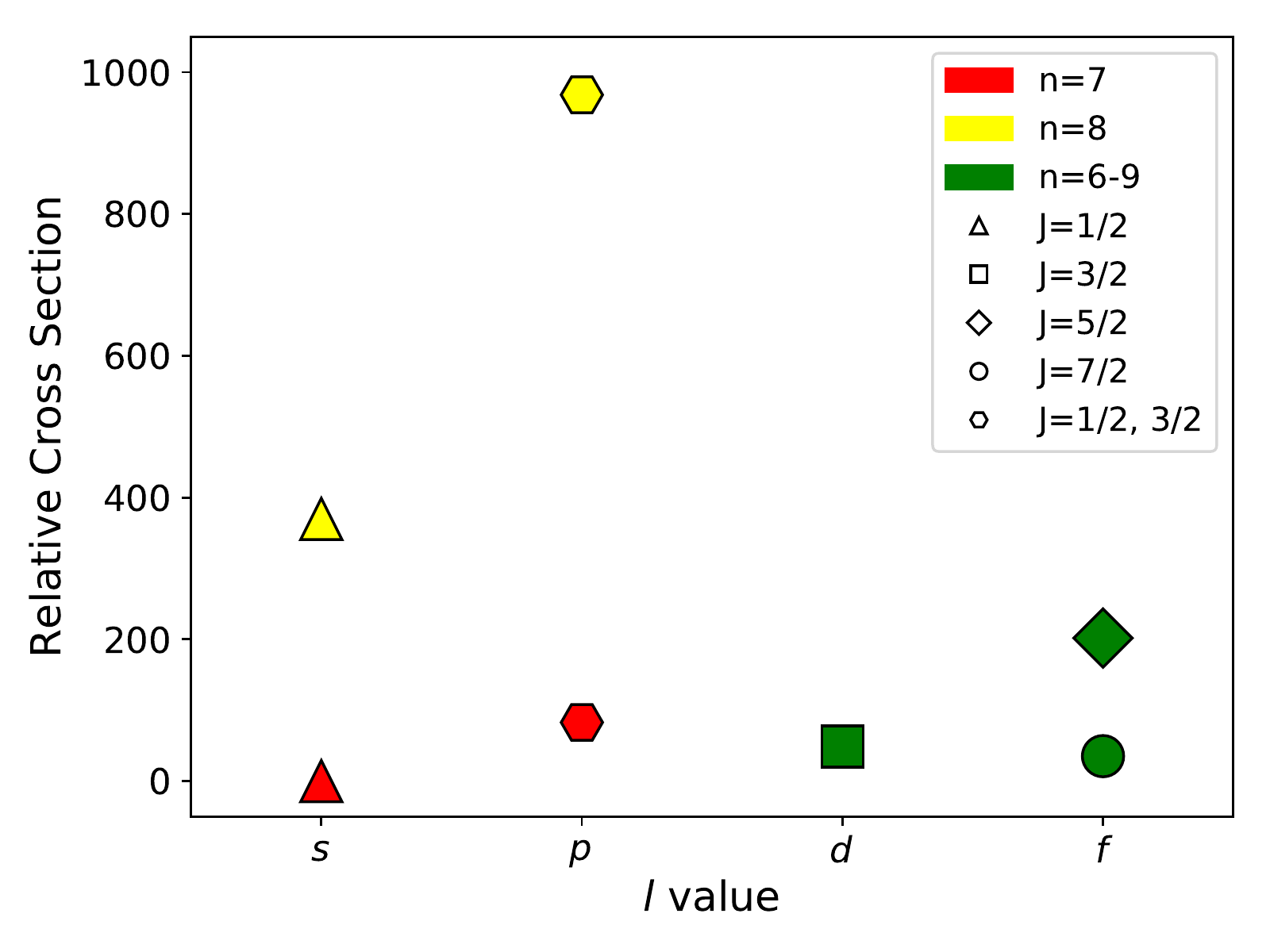}
 \end{center}
 \renewcommand{\baselinestretch}{1}
\small\normalsize
\begin{quote}
\caption[]{Relative $n, l, j, J$-resolved capture cross sections for H-like S+He CX. Each point reflects one quantum state of electron capture; the color, symbol, and x-value of each point describes the quantum state presented. This shows that the primary capture channel is into $8p$, followed in strength by $8s$.}
\label{fig:H_cs}
\end{quote}
\end{figure}

\begin{figure}
 \begin{center}
 \includegraphics[scale=0.5]{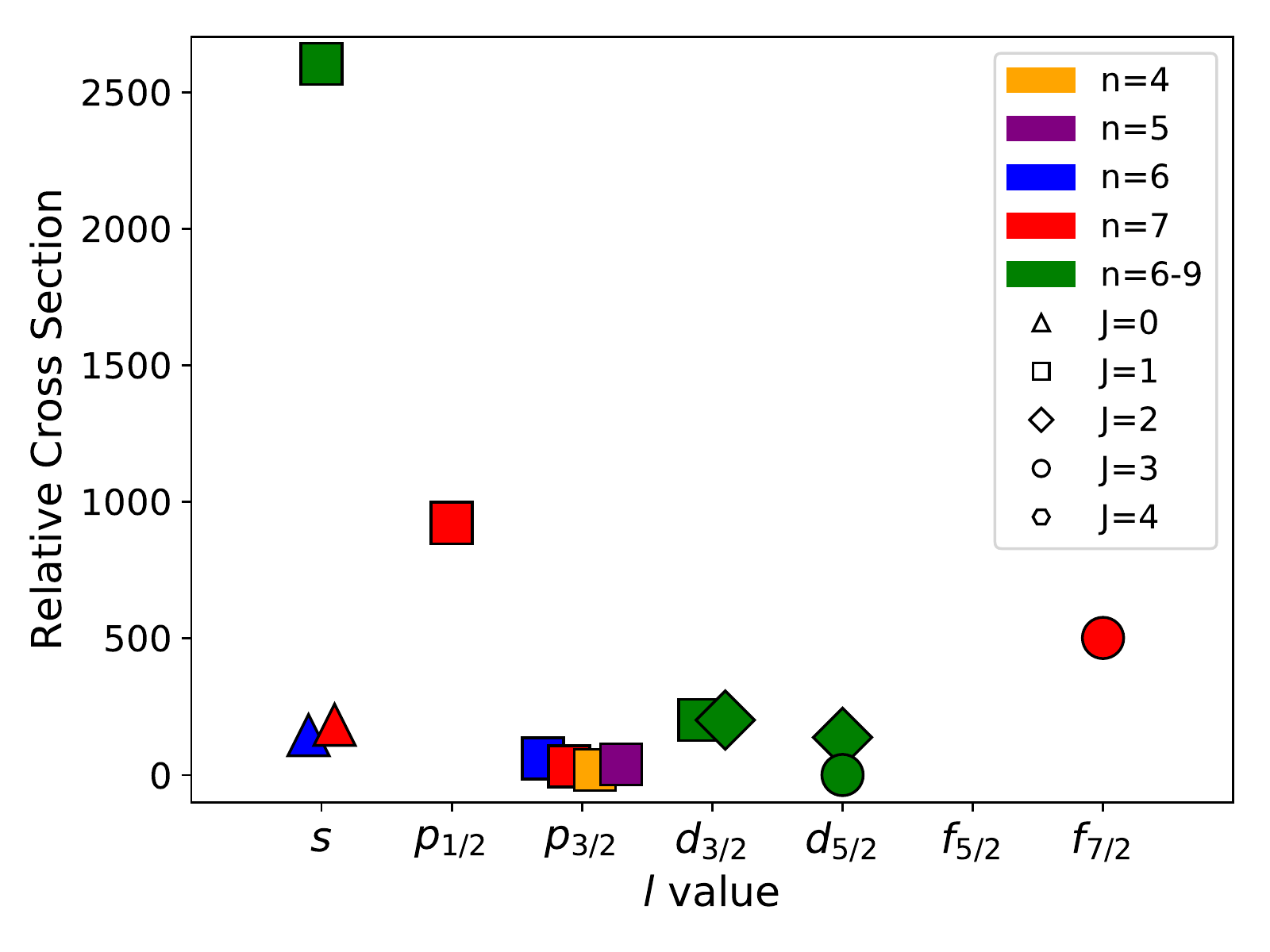}
 \end{center}
 \renewcommand{\baselinestretch}{1}
\small\normalsize
\begin{quote}
\caption[]{As in Figure \ref{fig:H_cs}, but for He-like S+He CX. We deduce that electrons are primarily captured into $7s (J=1)$ (see text).}
\label{fig:He_cs}
\end{quote}
\end{figure}

The relative cross section for each capture state described in our basis set is plotted in Figures \ref{fig:H_cs} and \ref{fig:He_cs}. Our results suggest that for H-like S in our experiments, the primary capture channel is into $8p_{1/2}$ and $8p_{3/2}$, which have nearly indistinguishable spectra, and thus were grouped together. Also significant is capture into $8s_{1/2}$. This is to be expected, since capture into low $l$ states should be preferred for low collision velocity EBIT CX experiments \citep{2000PhRvL..85.5090B}. For the He-like ion, capture into low $l$ states is also preferred, though in this case, the relative strength of capture into $s$ is higher than that of $p$. Although the spectra resulting from capture into $l=s$, $n=6-9$ are fairly indistinguishable, as capture into $n=7$ is preferred for $l=p$ and $f$, we assume that the point corresponding to the largest cross section at $l=s$ is also dominated by $n=7$. A fairly surprising result is the significance of capture into $nf$ for both ion species. It is possible that this results from double electron capture with at least one electron in $np$, followed by autoionization and a simultaneous transition of the second electron into $nf$.

\graphicspath{{images/}}

\graphicspath{{images/}}

\subsection{Comparison to MCLZ Calculations}

\begin{figure}
 \begin{center}
 \includegraphics[scale=0.5]{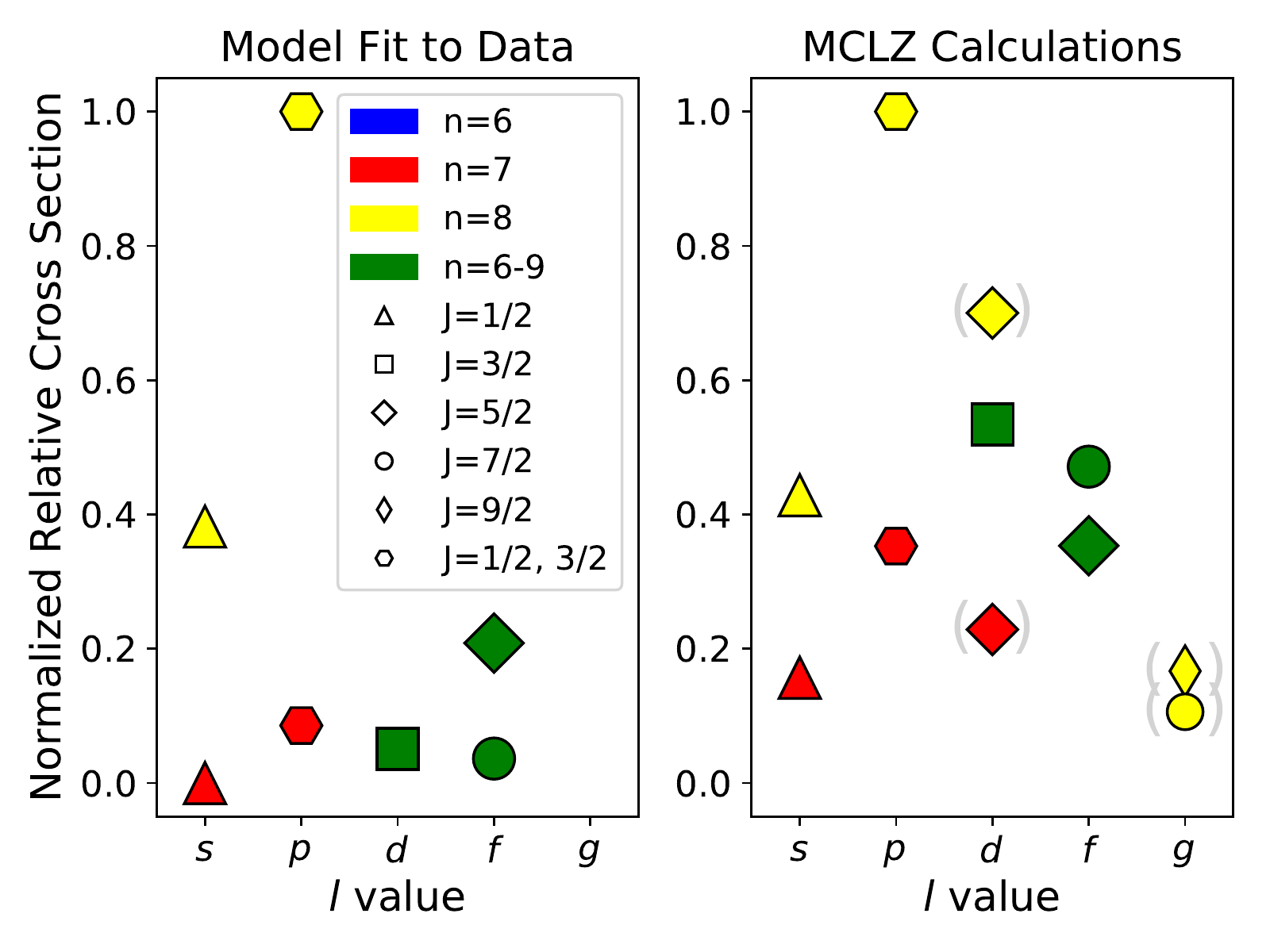}
 \end{center}
 \renewcommand{\baselinestretch}{1}
\small\normalsize
\begin{quote}
\caption[]{Normalized relative cross sections as obtained with the method described in this paper (left) and MCLZ calculations (right) for CX between H-like S and neutral He. Symbols are as in Figures \ref{fig:H_cs} and \ref{fig:He_cs}. To interpret this figure, one should compare the y-position of points of the same x-value, color, and symbol shape across the left-hand and right-hand plots. Points in parentheses refer to states that were not included in the original model fit presented here, but that were assigned a high relative cross section with the MCLZ calculations.}
\label{fig:model_compare}
\end{quote}
\end{figure}

To compare the cross sections obtained with our method, the Multi-Channel Landau-Zener method \citep{2016ApJS..224...31M} was used to calculate cross sections for CX between H-like S and neutral He. A low-energy $l$ distribution was applied to obtain $nl$-resolved cross sections at a collision energy of 10 eV/amu. This method was performed assuming LS-coupling, so the levels were then transcribed to their equivalent $jj$-coupled states using statistical $J$-weighting in order to compare to our model. Figure \ref{fig:model_compare} shows the most important capture states from the MCLZ calculations compared to our method. The cross sections obtained from these two methods show fair agreement. Both methods find that most capture is into $8s$ and $8p$ states, though the MCLZ calculation emphasizes capture into higher $l$ states ($d, f, g$) than our method does.

\graphicspath{{images/}}

\begin{figure}
 \begin{center}
 \includegraphics[scale=0.55]{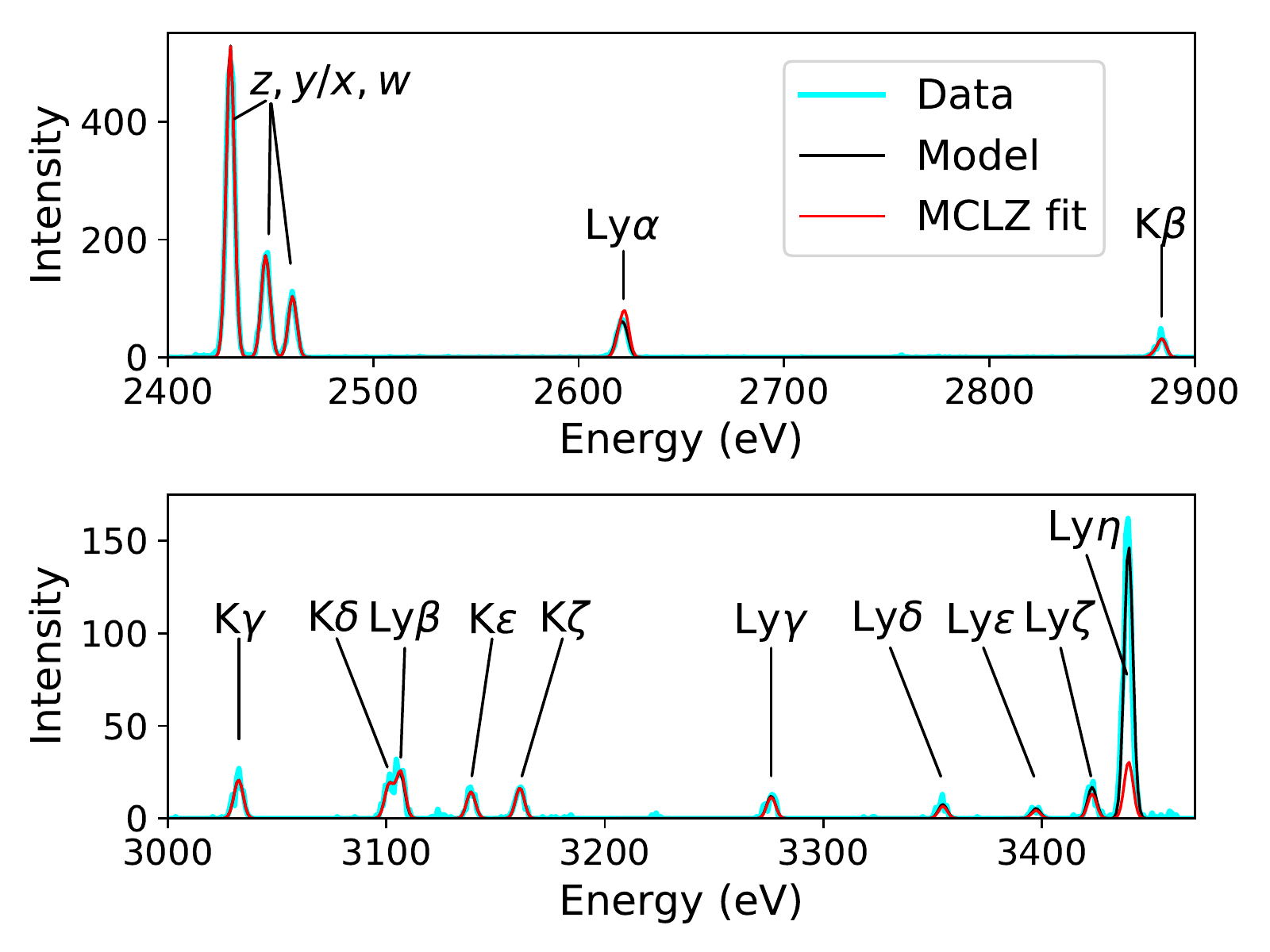}
 \end{center}
 \renewcommand{\baselinestretch}{1}
\small\normalsize
\begin{quote}
\caption{Data (cyan), model fit to the data through our fitting method (black), and simulated spectrum assuming MCLZ cross sections (red) to set the strength of the H-like lines; He-like lines are fit with our method. The strength of the H-like Lyman-$\eta$ line, as simulated with the MCLZ cross sections, shows rather poor fit to the data.}
\label{fig:simspec}
\end{quote}
\end{figure}

We then simulated the spectrum predicted by the MCLZ cross sections. We allowed the He-like lines to be fit according to the elements in our spectral basis set, and fixed the relative contributions of the H-like spectral basis set elements according to their MCLZ-calculated cross section. This spectrum is shown in Figure \ref{fig:simspec}. It can be seen that the match between experiment and theory for the H-like lines is fair; however, the strong Lyman-$\eta$ line is greatly underpredicted. This shows that although the MCLZ calculations correctly assign the highest cross section to capture states that lead to this strong line ($8p$), it overemphasizes the relative importance of capture states into higher angular momentum states, in particular $d,f$, and $g$. The Lyman-$\zeta$ and Lyman-$\alpha$ lines in the MCLZ spectrum also show poorer matches to the data than our method. 

\graphicspath{{images/}}

\section{Summary}\label{sec4}

We have presented an analysis tool that allows us to directly probe the physical processes at work during a CX experiment by extracting state-selective relative cross sections. We can then directly and quantitatively compare key parameters of a CX collision between experiment and theory, such as MCLZ calculations for K-shell ions. In a subsequent paper, we will present progress on improvements to this modeling process, in particular, experimenting with using $n$-selective cross sections from the literature or experiments to initially constrain our considered excited states (step 2), breaking degeneracies that exist in our spectral basis set (step 3), randomizing the selection of the spectral basis set (step 3), and results from using different fit statistics (step 4), in addition to expanded MCLZ results.

While this method reproduces our experimental S$^{16+}$+He CX data extremely well, we have shown that a simulated spectrum resulting from the corresponding MCLZ calculations does not fit as well. This underscores the need for continued theoretical modeling and experimental benchmarking. In parallel with these continued efforts, this method allows us to identify the most important quantum states during electron transfer---the most important parameter for correctly simulating a CX spectrum---and can aid in determining more detailed diagnostics for CX in astrophysical observations. This combination of experiment and theory will bring us closer to understanding the detailed atomic physics of CX, properly identifying it in our astrophysical observations, and harnessing its diagnostic power. 

\bibliography{xmmbib}

\end{document}